\documentclass[revsymb,bibnotes,aip,apl,amsmath,amssymb,reprint,superscriptaddress,floatfix]{revtex4-1}

\usepackage[]{graphicx}
\usepackage{bm}
\usepackage{float}
\usepackage{lineno}
\usepackage{hyperref}
\usepackage[all]{hypcap}
\usepackage{chngcntr}

\usepackage{lipsum}
\usepackage{slashed}
\usepackage{amsmath}
\usepackage{amsfonts}
\usepackage{amssymb}%
\usepackage{balance}

\usepackage{mhchem}

\makeatletter
\newcommand{\colorcaption}[2][]{%
  \begingroup%
  \renewcommand{\@caption@fignum@sep}{ (Color online). }%
  \caption[#1]{#2}%
  \endgroup%
}
\makeatother

\begin{document}

\author{T. Hache}
\affiliation{Helmholtz-Zentrum Dresden--Rossendorf, Institute of Ion Beam Physics and Materials Research, Bautzner Landstra\ss e 400, 01328 Dresden, Germany}
\affiliation{Institut f\"ur Physik, Technische Universit\"at Chemnitz, 09107 Chemnitz, Germany}

\author{Y. Li}
\affiliation{Department of Physics and Astronomy, Johns Hopkins University, MD 21218 Baltimore, USA}

\author{T. Weinhold}
\affiliation{Helmholtz-Zentrum Dresden--Rossendorf, Institute of Ion Beam Physics and Materials Research, Bautzner Landstra\ss e 400, 01328 Dresden, Germany}
\affiliation{Technische Universit\"at Dresden, 01062 Dresden, Germany}

\author{B. Scheumann}
\affiliation{Helmholtz-Zentrum Dresden--Rossendorf, Institute of Ion Beam Physics and Materials Research, Bautzner Landstra\ss e 400, 01328 Dresden, Germany}

\author{F. J. T. Gon\c{c}alves}
\affiliation{Helmholtz-Zentrum Dresden--Rossendorf, Institute of Ion Beam Physics and Materials Research, Bautzner Landstra\ss e 400, 01328 Dresden, Germany}

\author{O. Hellwig}
\affiliation{Helmholtz-Zentrum Dresden--Rossendorf, Institute of Ion Beam Physics and Materials Research, Bautzner Landstra\ss e 400, 01328 Dresden, Germany}
\affiliation{Institut f\"ur Physik, Technische Universit\"at Chemnitz, 09107 Chemnitz, Germany}

\author{J. Fassbender}
\affiliation{Helmholtz-Zentrum Dresden--Rossendorf, Institute of Ion Beam Physics and Materials Research, Bautzner Landstra\ss e 400, 01328 Dresden, Germany}
\affiliation{Technische Universit\"at Dresden, 01062 Dresden, Germany}

\author{H. Schultheiss}
\affiliation{Helmholtz-Zentrum Dresden--Rossendorf, Institute of Ion Beam Physics and Materials Research, Bautzner Landstra\ss e 400, 01328 Dresden, Germany}
\affiliation{Technische Universit\"at Dresden, 01062 Dresden, Germany}

\title{Bipolar spin Hall nano-oscillators}

\begin{abstract}
We demonstrate a novel type of spin Hall nano-oscillator (SHNO) that allows for efficient tuning of magnetic auto-oscillations over an extended range of gigahertz frequencies, using bipolar direct currents at constant magnetic fields. This is achieved by stacking two distinct ferromagnetic layers with a platinum interlayer. In this device, the orientation of the spin polarised electrons accumulated at the top and bottom interfaces of the platinum layer is switched upon changing the polarity of the direct current. As a result, the effective anti-damping required to drive large amplitude auto-oscillations can appear either at the top or bottom magnetic layer. Tuning of the auto-oscillation frequencies by several gigahertz can be obtained by combining two materials with sufficiently different saturation magnetization. Here we show that the combination of NiFe and CoFeB can result in 3~GHz shifts in the auto-oscillation frequencies. Bipolar SHNOs as such may bring enhanced synchronisation capabilities to neuromorphic computing applications.

\end{abstract}   
\maketitle
\maketitle

Spin Hall nano-oscillators (SHNOs) convert a direct current input into a microwave signal making them promising candidates for technologies such as miniaturized microwave generators or spin wave emitters \cite{Divinskiy2018} as well as artificial neurons in neuromorphic computing systems\cite{Zahedinejad2020}.\par 
Conventional SHNOs are comprised of a magnetic thin film adjacent to a non-magnetic heavy metal thin film\cite{Demidov2014}. The underlying mechanisms are the spin Hall effect (SHE), which contributes to the emergence of a spin polarised current in the heavy metal, and spin transfer torque (STT)\cite{Slonczewski1996, Berger1996}, which plays the role of inducing the large amplitude auto-oscillations (AOs) in the magnetic layer, provided the current densities are sufficiently high. \par 
High current densities can be achieved in laterally structured thin films, such as nano-constrictions\cite{Demidov2014,Zahedinejad2020}, close contact needles\cite{Demidov2012} or the apexes of bent magnetic stripes\cite{Sato2019}. The AO frequencies and the spectral weight depend on the demagnetization field landscape of a given nano-structure, material parameters and on external parameters such as magnetic field, \textit{dc} current and alternating currents. Indeed, the external magnetic field allows the tuning of the AO frequencies by several gigahertz\cite{Dvornik2018}, while other ways, such as varying the \textit{dc} current or injection locking only cause frequency shifts in the order of hundreds of megahertz\cite{Ulrichs2014,Wagner2018,Hache2019,Hache2020}. However, from a device perspective it is often impractical or difficult to change the external magnetic field conditions. Therefore, a different approach is necessary in order to achieve tunable broadband frequency operation of SHNOs. \par
 In this Letter, we present an approach that enables tunable broadband frequency operation without having to change the external magnetic field. This is achieved by incorporating a second magnetic layer to an SHNO. The now called double-layer SHNOs exhibit gigahertz shifts in the AO frequency upon switching the polarity of the \textit{dc} current. In single layer SHNOs the switching of the polarity does not induce AOs due to the increased damping caused by a change in the direction of the STT, while in the double-layer SHNOs, the switching of the current polarity means that the AOs are induced in the second magnetic layer.\par
\begin{figure}[t!]
\begin{center}
\scalebox{1}{\includegraphics[width=8.5 cm, clip]{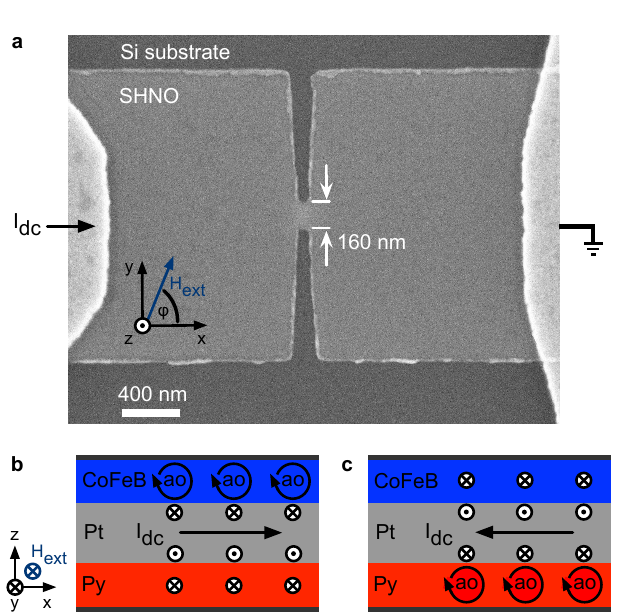}}
\colorcaption{\label{fig1}(a) SEM image of the constriction-based SHNO and an overlay of the external magnetic field and \textit{dc} current directions. (b)-(c) Schematic illustrations of the SHNO layer-stack. Both magnetic films are magnetised in the $y$-direction by an external magnetic field, $H_{ext}$. The arrows on the Pt layer pointing in the $y$-direction illustrate the spin accumulation due to the SHE provided a \textit{dc} current ($I_{dc}$) flows in the $x$-direction. In (b), the AOs are activated on the CoFeB layer while in (c) the AOs are activated in the NiFe following a change in the polarity of $I_{dc}$.}

\end{center}
\end{figure}

 
Figure~\ref{fig1}(a) shows a top-view scanning electron microscopy (SEM) image of the constriction-based SHNO geometry under investigation. The SHNOs were fabricated by means of electron-beam-lithography (EBL), followed by sputtering of the various thin film layers and subsequent lift-off. The fabrication process was applied on two different layer stacks: layer stack (A) consisting of, from top to bottom, \ce{Co40Fe40B20}(5~nm)/\ce{Pt}(7~nm)/\ce{Ni81Fe19}(5~nm); and layer stack (B), in which the position of the magnetic layers is flipped. The capping and the seed layers consist of 2~nm thick Tantalum. The electrical contacts were patterned using an additional fabrication step followed by thermal evaporation of the metals (Cr/Au) and lift-off.\par 

Figures~\ref{fig1}(b) and (c) illustrate the \textit{dc} current and external magnetic field conditions necessary for the generation of spin Hall driven AOs, which are consistently used throughout the manuscript. An in-plane magnetic field is applied in the \textit{$+$y}-direction in order to align the magnetization perpendicular to the \textit{dc} current. In this geometry the STT efficiency is maximized\cite{Slonczewski1996, Berger1996}. \par In a heavy metal with positive spin Hall angle, such as Pt\cite{Hoffmann2013,Sinova2015} , a \textit{dc} current flowing in the \textit{$+$x} direction leads to accumulation of positive and negative spin polarised electrons in the \textit{y} direction at the top and bottom interfaces, respectively, due to the SHE\cite{Dyakonov1971, Hirsch1999}. The spin current will act on the adjacent magnetic layers via the STT mechanism\cite{Slonczewski1996, Berger1996,Liu2012}. On the top interface, the polarised spins are parallel to the magnetic moment, so AOs can be achieved due to appropriate compensation of the damping term\cite{Slavin2009} by increasing the current density in the nano-constriction\cite{Dvornik2018}. On the other hand, no AOs are expected in the bottom layer due to increased damping originating from anti-parallel alignment between the polarised spin current and the magnetic moment. Thus, AOs in the bottom layer can only be achieved when reversing the DC current whilst maintaining the same magnetic field polarity.\par
Since the magnetic layers employed in this stack have considerably different saturation magnetization (\ce{Co40Fe40B20}: 1080 kA/m; \ce{Ni81Fe19}: 640 kA/m, obtained from VSM measurements), one expects that the AO frequencies of both materials will differ accordingly. Therefore the microwave output characteristics of double layer SHNOs exist over a broad range of frequencies and can be controlled by reversing the direction of the \textit{dc} current. \par

Prior to discussing the performance of the double-layer SHNO it is important to establish an understanding of the AO behaviour of both CoFeB and NiFe which are the top layers of stack (A) and (B), respectively. For this purpose, micro-focused BLS\cite{Sebastian2015} was used to measure the frequency and spectral weight of the AOs. Micro-focused BLS is a particularly sensitive technique because even thermal spin waves can be detected. This particular apparatus includes electrical contacts, enabling us to apply a \textit{dc} current perpendicular to the magnetic moment ($\mathrm{\varphi}$=~90~$\deg$), which is the angle of highest spin transfer torque efficiency\citep{Slonczewski1996, Berger1996}. This way, we were able to measure the \textit{dc} current threshold, above which large amplitude AOs emerge. We stress the importance of clearly defining the polarity of both the magnetic field and the \textit{dc} current as the four possible combinations will be discussed later in the manuscript.\par

\begin{figure}[]
\begin{center}
\scalebox{1}{\includegraphics[width=8.5 cm, clip]{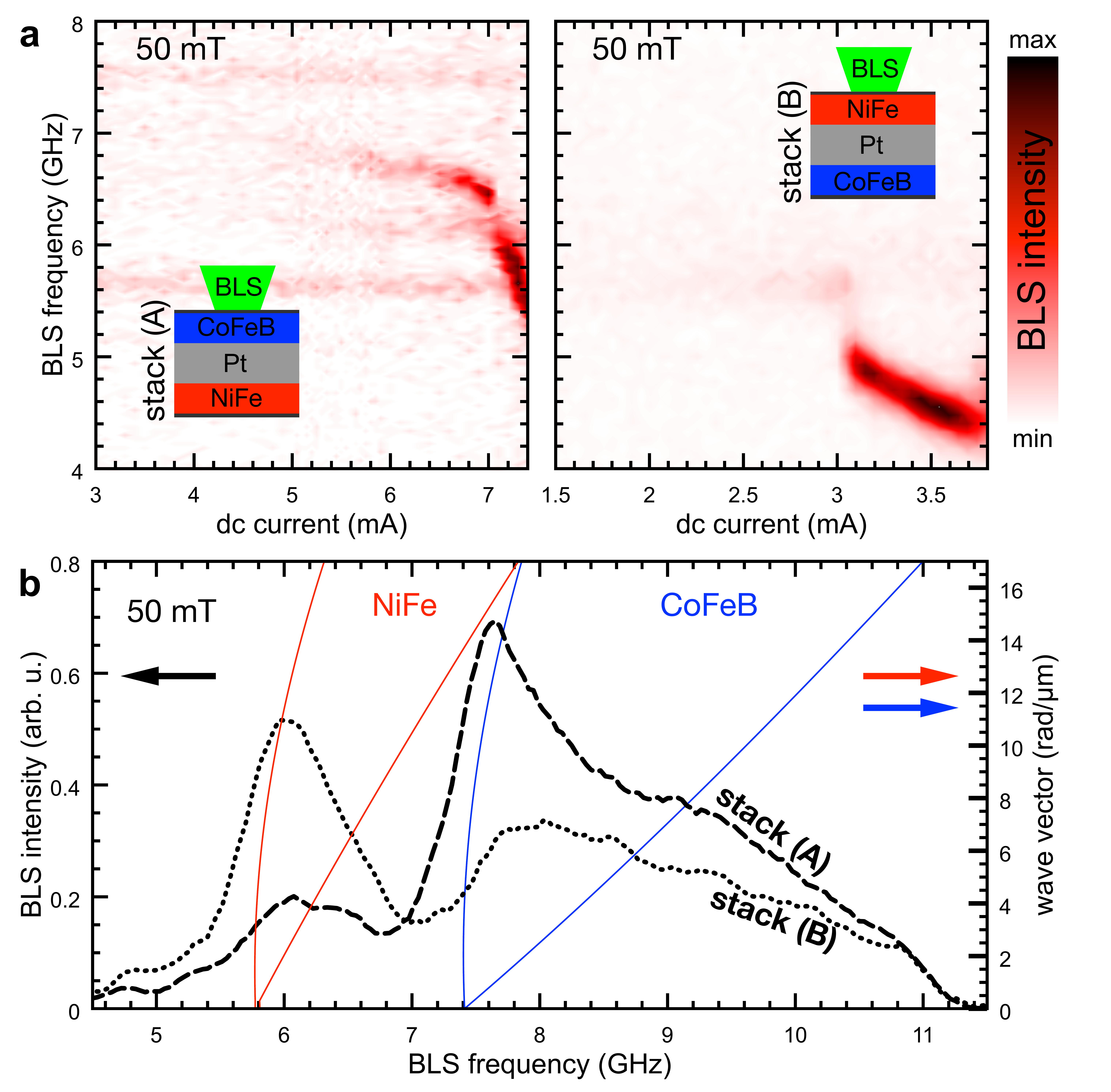}}
\colorcaption{\label{fig2} (a) BLS intensity plotted as a function of frequency and \textit{dc} current for stack (A) and its reversed layer stack (stack (B)), measured at $+$50~mT. These plots highlight the different \textit{dc} current threshold and frequency range of the two layers. (b) (axis on the left) thermal BLS intensity of the unstructured continuous thin films of stacks (A) and (B), measured at $+$50~mT; (axis on the right) wavevector plotted as a function of frequency for NiFe (5~nm) and CoFeB (5~nm).}
\end{center}
\end{figure}

Figure~\ref{fig2}(a) shows intensity maps of the AOs as a function of the BLS frequency and \textit{dc} current for stacks (A) and (B), measured at $+$50~mT. In stack~(A), AOs emerge at 6.8 GHz at a current threshold of $+$6~mA. With increasing the \textit{dc} current to $+$7.4~mA one observes a decrease in the AO frequency to 5~GHz. Above $+$7.4~mA the device performance would be affected by Joule heating, causing irreversible damage to the SHNO, as observed in replicas of this device. In stack (B), the current threshold is about $+$3~mA with AOs emerging at 4.8~GHz. The frequency of the AOs decreases to 4.2~GHz as the current is increased to $+$3.8~mA. The decrease in AO frequency of both stacks is consistent with previous reports concerning in-plane magnetised SHNOs\cite{Demidov2014}.\par 
The results discussed with Fig.~\ref{fig2}(a) clearly demonstrate that the critical current density for AOs can be reached in these samples. Moreover, one can clearly distinguish between the two different AO frequencies due to the different saturation magnetization of both materials. \par

Figure~\ref{fig2}(b) shows the thermal BLS spectra measured in reference continuous thin films of stacks (A) and (B), at $+$50~mT. In stack (A) the dominant BLS intensity is centred at 7.6~GHz, while in stack (B) the dominant BLS intensity is centred at 6~GHz. Comparatively, the frequencies of the peaks measured in the thermal spectra are larger than those observed in the micro-BLS signal obtained in the nano-constriction shown in Fig.~\ref{fig2}(a). The reason is that the dispersion relation within the magnetic constriction is expected to yield lower frequencies primarily due to the larger demagnetization fields\cite{Demidov2014, Slavin2009}.

In order to confirm the origin of the thermal spectra we calculated the dispersion relation in 5~nm thick CoFeB (blue) and NiFe (red) films\footnote{Parameters used for the calculation of the dispersion relation: $M_s$, thickness and field as given; $\gamma$=~176~GHz/T; $A_{ex}$: (CoFeB)= 15~pJ/m\cite{Conca2013}; NiFe= 12.4~pJ/m\cite{Hrabec2016}.}. These results are plotted in Fig.~\ref{fig2}(b) using the vertical axis on the right hand side, whose wavevector range is coincident with that of a typical micro-BLS experiment\cite{Sebastian2015} (0 - 17~rad/um, objective lens with an NA of 0.75). Clearly the frequency region with higher BLS intensity of stack (A) is coincident with the calculated frequency range for CoFeB(blue lines). On the other hand, the thermal BLS spectrum of stack (B) is coincident with the bandwidth of the calculated dispersion relation for NiFe (red lines). \par
In addition, the thermal BLS spectra of both stacks have a secondary peak in the frequency range expected for the corresponding bottom layers. In the case of stack (A) the secondary peak coincident with the NiFe, while in stack (B) there is a secondary BLS peak at the frequency range expected for the CoFeB. The comparatively lower BLS intensity of the bottom layer is due to the finite penetration depth of the laser light through the various metallic layers\cite{Sebastian2015}. \par 
Moving forward, we will discuss the SHNO characteristics of both top and bottom layers of stack (A) using an all-electrical detection method. In this method, the measured quantity is the anomalous magneto-resistance (AMR) change due to AOs and is obtained via a microwave amplifier in combination with a spectrum analyzer. In order to obtain a significant change in the AMR due to AOs the external magnetic field is set to $\mathrm{\varphi}$=~68~$\deg$. With the choice of applied field angle one has to make a compromise between the magnetic field angle that yields higher STT efficiency (90~$\deg$) and the angle that results in a larger AMR signal (45~$\deg$). \par
\begin{figure}[]
\begin{center}
\scalebox{1}{\includegraphics[width=8.5 cm, clip]{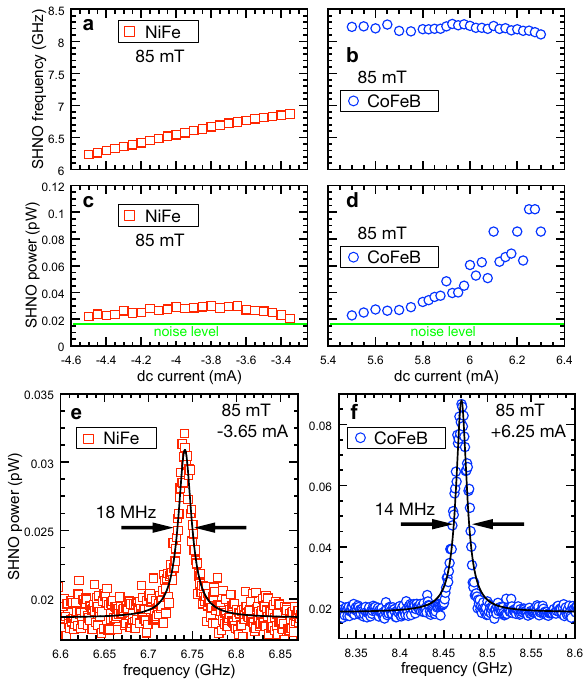}}
\colorcaption{\label{fig3} (a)-(d) AO frequency and power plotted as a function of the DC current at a fixed external magnetic field of $+$85~mT. The noise level (green base line) indicates the threshold for the detection of the microwave signal. (e)-(f) Examples of the raw AO signal measured at $+$85~mT for negative and positive DC current values and corresponding fits using Lorentzian lineshapes.}
\end{center}
\end{figure}
Figures~\ref{fig3}(a)-(d) show the frequency and amplitude of the AOs for positive (left) and negative (right) polarities of the \textit{dc} current, at a constant external magnetic field of $+$85 mT. At this magnetic field value, one observes good AO characteristics at both positive and negative \textit{dc} currents. At negative \textit{dc} current values, the AOs occur at lower frequencies and vary between 6.9~GHz, at $-$3.3~mA, and 6.2~GHz at $-$4.5~mA with a maximum generated microwave power of 0.032~pW\footnote{Within the range of the linear response of the microwave amplifier used for the detection of the AMR modulation of the resistance. The amplifier gain (averaged to 71~dB) was subtracted from the data in order to determine the true SHNO power.}, at about $-$3.65~mA (see Fig.~\ref{fig3}(e)). Having learned from Fig.~\ref{fig2}(a) that positive \textit{dc} currents and positive fields result in AOs originating from the CoFeB layer of stack (A), we can expect that negative \textit{dc} currents will trigger AOs originating from the NiFe layer. Note also that the comparatively lower frequencies and the \textit{dc} current threshold further confirm that the AOs are generated in the NiFe layer. \par 

The results obtained at positive \textit{dc} currents are consistent with the micro-BLS data shown in Fig.~\ref{fig2}(a) and are therefore attributed to the CoFeB layer. Within the \textit{dc} current range applied in these experiments, the AO frequency of the CoFeB varies only slightly at around 8.3~GHz with a tendency to generate larger AO powers (0.1~pW) for larger \textit{dc} current values ($+$6.25~mA).\par 

\begin{figure*}[t]
\begin{center}
\scalebox{1}{\includegraphics[width=17 cm, clip]{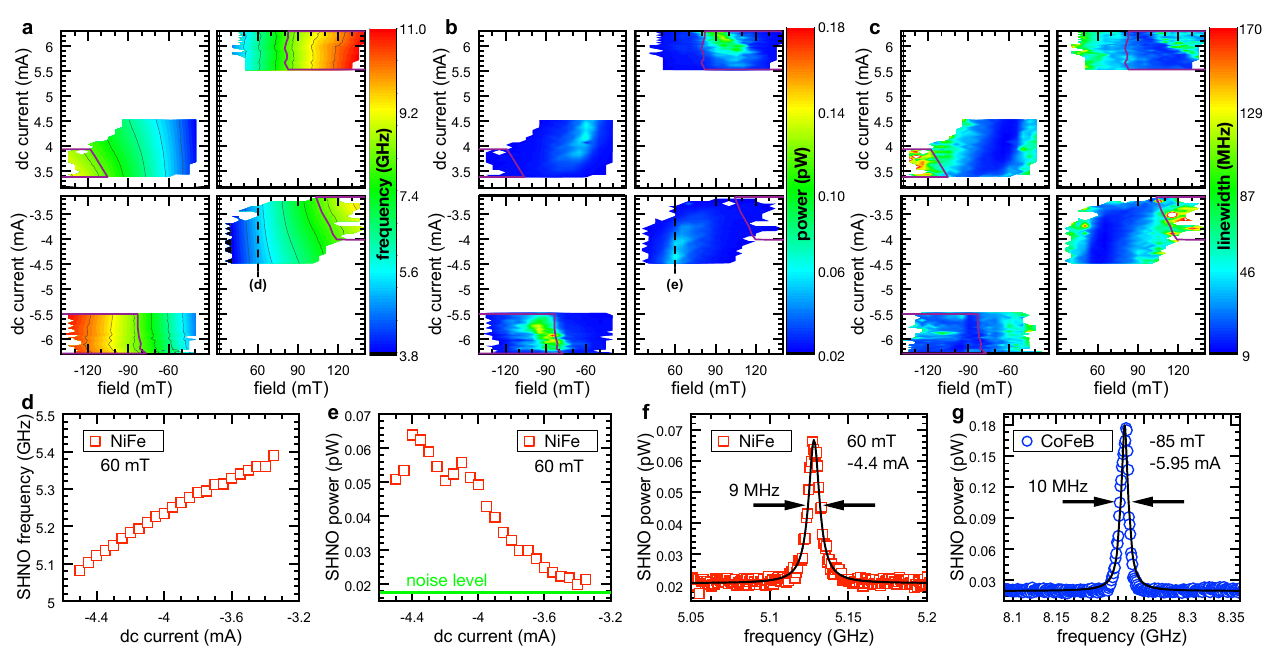}}
\colorcaption{\label{fig4} Colour plots of the generated AO frequency (a), power (b) and linewidth (c) plotted as a function of applied \textit{dc} current and external magnetic field. The plots shown here are a result of interpolating the fitted data. The regions in blank correspond to field/current values where AO signals are below the detection threshold. The regions delimited by the lines in violet correspond to the field/current values, whose AO frequency is above 8~GHz, where the response of the amplifier is not linear, resulting in larger uncertainty in measuring the power and the linewidth. (d)-(e) Line plots at 60~mT, where the NiFe layer shows large SHNO power. (f) and (g) show the ($I_{dc}$, $H_{ext}$) conditions that yield the largest amplitude and narrowest linewidth on the NiFe and CoFeB layers, respectively.}
\end{center}
\end{figure*}

The two spectra shown in Fig.~\ref{fig3}(e)-(f) correspond to the AO signals with higher power obtained for both polarities of the \textit{dc} current using a fixed magnetic field. These results demonstrate that switching of the AO frequencies in the GHz range in a double-layer SHNO at a fixed magnetic field can be achieved by means of changing the polarity of \textit{dc} current. \par

In order to demonstrate the broadband frequency tunability of these devices, we measured the AO characteristics over a wide range of field and \textit{dc} current values. Figure~\ref{fig4} (a)-(c) shows three panels summarising the AO frequency (a), generated SHNO power (b) and frequency bandwidth (c) as a function of the \textit{dc} current and the external magnetic field where AOs were observed. The plotted quantities were obtained following a fitting of the measured raw spectra with Lorentzian lineshapes. \par 
Of interest is the immediate knowledge that the AO frequencies can vary between 3.8 and 11~GHz, depending on the choice of ($I_{dc}$, $H_{ext}$) values. Similarly, in Figs.~\ref{fig4}(b)-(c) one can point out the ($I_{dc}$, $H_{ext}$) values which yield larger AO output power with narrower linewidths.\par 

Figures~\ref{fig4}(d)-(e) show the ($I_{dc}$, $+$60~mT) line traces of the AO frequency and power corresponding to the NiFe layer. This magnetic field value yields the largest AO power originating from the bottom layer (NiFe), particularly at higher \textit{dc} current values, where the largest SHNO power (0.07~pW) together with narrowest linewidth (9~MHz) were obtained ($-$4.4~mA, $+$60~mT). This line trace is plotted in Fig.~\ref{fig4}(f).\par
The best performance obtained from the CoFeB layer was observed at ($-$5.95~mA, $-$85~mT), as shown in Fig.~\ref{fig4}(g). Here the SHNO power obtained is 0.18~pW with a linewidth of 10~MHz. 

Importantly, the summary provided here highlights the advantages of the double-layer SHNO compared to a single layer structure. Firstly, we demonstrate the ability to generate AO signals across the four quadrants of the \textit{dc} current versus magnetic field diagrams shown above. Secondly, it opens the possibility to generate AOs over a broader frequency range by using bipolar \textit{dc} currents at a constant external magnetic field. \par 

Our findings open many experimental possibilities in the field of multilayer SHNOs. Auto-oscillation frequencies can be tuned for specific experimental requirements by combining different materials (with different saturation magnetization). With appropriate choice of the materials, one could also achieve similar functionality by changing the magnetic field polarity at constant \textit{dc} currents. In addition, it would be beneficial to use materials with high resistance in order to limit the \textit{dc} current flow to the Pt layer. We anticipate that very promising results in terms of frequency tunability can be achieved when combining insulators (YIG, low $\mathrm{M_s}$) with amorphous (high resistance metals (CoFeB, high $\mathrm{M_s}$)). For instance, one may consider using this bipolar SHNO to obtain two synchronisation frequencies in the same nano-constriction, which could act as a neuron in a neuromorphic device. Other steps could be taken, for instance the introduction of thin films with strong anisotropies for magnetic field free SHNO operation or the use of domain walls and other non-collinear spin textures.\par
 
Financial support by the Deutsche Forschungsgemeinschaft is gratefully acknowledged within program SCHU2922/1-1. Lithography was done at the Nanofabrication Facilities (NanoFaRo) at the Institute of Ion Beam Physics and Materials Research at HZDR. 

\end{document}